\documentclass[twocolumn,prl,aps,showpacs,floatfix]{revtex4}

\usepackage{graphicx}%
\usepackage{dcolumn}
\usepackage{amsmath}

\newcommand{\ice}[1]{\relax}

\newcommand{\as}{a_s}

\newcommand{\als}{\alpha_s}
\newcommand{\beq}{\begin{equation}}
\newcommand{\eeq}{\end{equation}}
\newcommand{\bea}{\begin{eqnarray}}
\newcommand{\eea}{\end{eqnarray}}
\newcommand{\g}{\gamma}

\newcommand{\asthree}{{}^{2\times{\cal O}(\as^3)}_{{\rm no}\;{\cal O}(\as^3)}}
\newcommand{\asfour}{{}^{2\times{\cal O}(\as^4)}_{{\rm no}\;{\cal O}(\as^4)}}
\begin{document}

\title{
{
 \vspace*{-14mm}                                                            
                                                                        
\centerline{\normalsize\hfill SFB/CPP-04-72}                                       
\centerline{\normalsize\hfill TTP04-28}                                        
\centerline{\normalsize\hfill hep-ph/0412350}                                   
{}} 
\vspace{3mm}
Strange Quark Mass from Tau Lepton Decays with 
${\cal O}(\alpha_s^3)$ Accuracy
\vskip.3cm
     }
\author{P.~A.~Baikov}
\affiliation{Institute of Nuclear Physics, Moscow State University,
Moscow~119899, Russia
        }
\author{K.~G.~Chetyrkin\thanks{{\small Permanent address:
Institute for Nuclear Research, Russian Academy of Sciences,
 Moscow 117312, Russia}}}

\author{J.~H.~K\"uhn}
\affiliation{Institut f\"ur Theoretische Teilchenphysik,
  Universit\"at Karlsruhe, D-76128 Karlsruhe, Germany}

\begin{abstract}
\vspace*{.7cm}
\noindent
The first complete  calculation of the quadratic
quark mass correction to the correlator of the two  currents
relevant for the strangeness-changing  semihadronic tau-decay rate 
is presented {\em including its real} part at the four-loop level. 
This allows to perform   the
extraction of the strange quark mass $m_s$ from the decay width of the
tau-lepton with  {\em full} ${\cal O}(\alpha_s^3)$ accuracy. 
In  agreement with previous estimates the newly computed $\alpha_s^3$
term proves to be rather large. This justifies inclusion of the
similarly estimated $\als^4$ term in phenomenological
analysis. Combined with an updated value  of $V_{us}$ and an
``'improved'' version of the renormalization group (RG)  improvement of the
perturbative series this leads to an increase of the central value of
$m_s$ by about 20\% and a partial reduction of the theoretical
uncertainty by about 50\%.

\end{abstract}


\maketitle

\section{Introduction}   
    
The investigation of Cabbibo suppressed semileptonic $\tau$ decays has
developed into one of the important themes of $\tau$-lepton
physics. Combined theoretical and experimental studies provide an
independent and fairly precise value for the strange quark mass $m_s$
and may in the future also lead to a competitive determination of the
Cabbibo-Kabayashi-Maskawa matrix element $V_{us}$ 
\cite{Barate:1999hj,Abbiendi:2004xa,Chetyrkin:1993hi,Maltman:1998qz,Chetyrkin:1998ej,Pich:1998yn,%
Korner:2000wd,Chen:2001qf,Maltman:2002wb}. 
The decay proceeds into vector and axial current induced final states,
which can be further separated experimentally into the spin zero and
spin one components \cite{Kuhn:1992nz}. In addition to the total rate also various
moments of the distribution in $s$, the invariant mass of the hadronic
state, can be considered and will be discussed below.

The theory prediction for the rate and moments is based on
perturbative QCD with small contributions from nonperturbative
condensates
\cite{Braaten:1991qm}. 
The
calculation is greatly simplified by the smallness of the strange
quark mass which justifies an expansion in powers of
$m_s^2/M_{\tau}^2$. Up to now the $m_s$ dependence of the total rate 
\ice{and a specific combination of moments $(L+T)$ which is more stable
against inclusion of higher orders, 
}
was only known up to order
$\als^2$. To push the precision further, estimates for the third order
coefficient were used which were based on the principle of minimal
sensitivity
\cite{Stevenson:1981vj} (PMS) or fastest apparent convergence 
\cite{Grunberg:1984fw}  (FAC). 
In the present paper the exact result for this coefficient is
presented. It is deduced  from the first complete calculation of the finite
part of a specific correlator in four loop approximation. This
calculation is based on the conceptual developments described in
\cite{Baikov:tadpoles:96,BkvSthr:tadples:98,Baikov:criterion:00}
and requires extensive use of computer algebra \cite{Vermaseren:2000nd}.  
The result confirms the
PMS/FAC estimate and justifies the use the same approach for an
estimate of the $\als^4$ coefficient. Inclusion of this next term
leads to a decrease  of the central value of $m_s$ by a few MeV
and a partial reduction of the theoretical uncertainty by 50\%.

The phenomenological analysis  will be based on the, by now standard,  approach  
of contour improved perturbation theory (CIPT)
\cite{Pivovarov:1991bi,LeDiberder:1992te}. 
Two variants  will be discussed. We will argue in
favour of one of them  which leads in total to more stable results.
The newest developments for $V_{us}$ will be included
in the analysis.

The main outcome of the study of the total hadronic $\tau$ width
(normalized to the known leptonic width)
\begin{equation}
R_\tau=\frac{\Gamma(\tau\rightarrow{\rm hadrons}
  +\nu_\tau)}{\Gamma(\tau\rightarrow l+\bar\nu_l+\nu_\tau)}
\label{Rtau1}
\end{equation}
has been a remarkable confirmation of perturbative QCD. It was found that
the value of the strong coupling constant $\alpha_s$ as obtained from
$R_{\tau}$ is in good agreement with  those obtained from completely
different experiments such as the  $Z$ boson decay  into hadrons
\cite{Barate:1998uf,Ackerstaff:1998yj,Chetyrkin:1996ia,EWWG:2003_v3}.
Furthermore,
the strangeness changing (Cabbibo-suppressed) part $R_S$ 
of the decomposition of the decay
rate and the moments  \cite{LeDiberder:1992te}
\bea                                                                   
\label{Rtaukl}                                                                     
R_\tau^{kl} &\equiv& \!\int\limits_0^{M_\tau^2}\! ds\, \biggl( 1 -                   
\frac{s}{M_\tau^2} \biggr)^{\!k}\!\biggl(\frac{s}{M_\tau^2}\biggr)^{\!l}           
\frac{d R_\tau}{ds} = R_{\tau,NS}^{kl} + R_{\tau,S}^{kl} \,,                       
\nonumber
\\
R_\tau^{00} &\equiv& R_\tau, \ \ R_{\tau,NS}^{00} \equiv R_{\tau,NS},
\ \ R_{\tau,S}^{00} \equiv R_{\tau,S}
{}
\eea
into  non-strange (NS) and strange (S) components  
can be used to determine the strange quark mass, one of the
fundamental parameters of the Standard Model. The moments as
introduced above can be experimentally determined from the measured
distribution in the invariant mass of the final state hadrons. Hadronic physics is 
encoded in  the quantities  $r^{kl}_{ij}$ defined through 
\begin{equation}
R_{\tau,NS}  + R_{\tau,S}= 3\, S_{\rm EW}
  \left(|V_{ud}|^2 \, r_{ud} +|V_{us}|^2 \, r_{us} 
\right)
{},
\label{Rtau2}
\end{equation}
where $v_{ij}$ stands for the CKM matrix and 
$S_{EW}$ for the universal electroweak
correction \cite{Marciano:1988vm,Braaten:1990ef}.  
The functions $r^{kl}_{ij}$ are directly related 
to the correlator of the charged current
$j_\mu(x) =\bar{\psi_i} \g_\mu(1-\g_5)\psi_j $
\bea
 i \int \mathrm{d} x\,  e^{iqx} 
\langle  
T\left[
j_\mu(x) j^{\dagger}_\nu(0)
\right]
\rangle
&=&
\nonumber
\\
(-g_{\mu\nu} q^2 +  q_{\mu}q_{\nu}) \Pi^{(q)}_{ij}(q^2)
      &+&  g_{\mu\nu} q^2 \Pi^{(0)}_{ij}
 \label{corr:def}
{}.
\eea
through 
\bea
\lefteqn{r_{ij}^{k,l} =  r_{ij}^{(q)k,l} +  r_{ij}^{(0)k,l} =} 
\label{rij}
\\
&{}&
2i\pi 
{\displaystyle
\oint_{|s|=M_{\tau}^2}
}
\frac{ds}{M_{\tau}^2}
\left[
w^{(q)k,l}(x)
\Pi^{(q)}_{ij}(s)
+
w^{(0)k,l}(x)
\Pi^{(0)}_{ij}(s)\right]
{}.
\nonumber
\eea
where $x=s/M^2_{\tau}$, and the weight functions $w^{(q)k,l} =
(1+2x)(1-x)^{k+2}\,x^l$, $w^{(0)k,l} = -2x^{l+1}(1-x)^{k+2}$. 
The behaviour of the perturbative series for the part of the integral
arising from $\Pi^{(q)}$ is more stable than the one from $\Pi^{(0)}$
\footnote{
The combinations $(q)$ and $(0)$ are equivalent to $L+T$ and $L$ in the
notation  of ref.~\cite{Pich:1998yn}.}.
However, the latter can either be modelled theoretically and
phenomenologically on the basis of scalar and pseudoscalar  resonance physics
\cite{Maltman:1998qz,Kambor:2000dj,Gamiz:2002nu,Gamiz:2004ar},
or, being solely determined  by spin zero contribution in the $\tau$ decays,
determined experimentally through  the analysis of angular distribution of the 
$K\pi$ and $K\pi\pi$, thus separating spin zero 
and spin one contributions \cite{Kuhn:1992nz}. 
Restricting to scalar and pseudoscalar channels one finds
\bea
R^{kl(J=0)}_{\tau,S} &=& \int_0^{M_{\tau}^2} {d s} 
\left(1-\frac{s}{M_{\tau}^2}\right)^k 
\left(\frac{s}{M_{\tau}^2}\right)^l
\frac{d R^{J=0}}{d s} 
\nonumber
\\
&=&
 \frac{-3}{2}\,  S_{EW}\,\,|V_{us}|^2\,\ r^{(0)k,l-1}_{us} 
\label{J=0:(L):connection}
{}
\eea
A specific weighted integral over scalar spectral function thus
directly determines the moments $r^{(0 )kl}_{ij}$. (A closely related
discussion along these lines, which deals with $J=1$ and $J=0$
spectral functions and the resulting moments separately and   the
corresponding  predictions can be found in
\cite{Chetyrkin:1998ej}.)  Since $m_u, m_d \ll m_s$ and $m_s \ll M_\tau$,
the function $r_{us}^{k,l}$ is well described by setting $m_u=m_d=0$ and
keeping only the leading and quadratic contributions to the 
correlator~(\ref{corr:def}),
viz.
\bea
\Pi^{(q)}_{ij}(q^2,m_s) &=& \frac{3}{16\pi^2} \left(
\Pi^{(q)}_0(q^2) + \frac{m_s^2}{Q^2} \Pi^{(q)}_{2,ij}(q^2) 
\right)
\label{Pi:q}
\\
\Pi^{(0)}_{ij}(q^2,m_s) &=&
 \frac{3}{16\pi^2} \frac{m_s^2}{Q^2} \Pi^{(0)}_{2,ij}(q^2) 
\label{Pi:0}
{}.
\eea
Here $Q^2 = -q^2$.  In the following, we limit ourselves by the
perturbative contributions 
(for a recent discussion of power-suppressed
contributions, see \cite{Gorbunov:2004wy}).
The small non-perturbative terms, as well
as the $m_s^4$ terms will be effectively included in the
phenomenological analysis at the end. 
As a result one has a
convenient decomposition
\beq
r_{u
d} = r_0 + \delta_{ud}, \ \ \ r_{us} = r_0 + \delta_{us} \ \ \
{}.
\eeq
Let us in addition define the  difference
\begin{equation}                                                                
\label{delRtaukl}                                                               
\delta R_\tau^{kl} \equiv \frac{R_{\tau,NS}^{kl}}{|V_{ud}|^2} -                 
\frac{R_{\tau,S}^{kl}}{|V_{us}|^2} =                                            
3\, S_{{\rm EW}}
 \,  \delta r^{kl}
{}.                                                    
\end{equation}  
which is a useful  combination to probe the $SU(3)$ breaking
effects as 
$\delta r^{kl} \equiv \Big(\delta_{ud}^{kl} -\delta_{us}^{kl}\Big)$  
vanishes in the limit of  exact $SU(3)$ flavour
symmetry.  In the spirit of the decomposition (\ref{rij}) one similarly defines 
related quantities $\delta R_\tau^{(q)kl}$, $\delta R_\tau^{(0)kl}$, 
$\delta r^{(q)kl}$ and $\delta r_\tau^{(0)kl}$
such that
$
\delta R_\tau^{kl}  = \delta R_\tau^{(q)kl} + \delta R_\tau^{(0)kl}
\nonumber
$
,
$
\delta r^{kl}  = \delta r^{(q)kl} + \delta r^{(0)kl}
\nonumber
{}.
$

The currently known fixed order perturbative  predictions for $r_0$, $\delta_{us}$ and
$\delta_{ud}$ can be shortly summarized as follows
\cite{Gorishnii:1991vf,Chetyrkin:1993hi,Maltman:1998qz}:
\ice{
\footnote{displayed below
are so called Fixed Order Perturbation Theory results.  A procedure of
``improving'' of pQCD predictions by a partial resummation of higher
order terms (CIPT) will be discussed later.}.
}
\bea
r_0                  &=&  
\label{Rtau:FO}
\left(
1 + \, a_s+ 5.202 \, a^2_s+ 26.37 \, a_s^3 
    \right)
{},
\\
\delta_{us}          &=&  
\frac{m_s^2}{M_{\tau}^2}
\left(
1 + 5.33\, a_s+ 46.0\, a^2_s 
    \right)
{},
\label{delta_us:FO}
\\
\delta_{ud}           &=&  
-0.35\  a^2_s \ 
\frac{m_s^2}{M_{\tau}^2}
\label{delta_ud:FO}
{},
\eea
where $a_s = \als(M_\tau)/\pi$.  Now, with $a_s \approx .1$ one
observes that the ``apparent convergency'' of the  series is
acceptable for $r_0$ but should be considered  at best as marginal for
$\delta_{us}$.  Clearly, the next term in (\ref{delta_us:FO}) it is
important for the interpretation of the measurements. Partial
results for the $\as^4$ term in eq.~(\ref{Rtau:FO}) and the $\as^3$
term in eq.~(\ref{delta_us:FO}) have been published
\cite{ChBK:tau:as4nf2}. We give the results of the
calculation of the missing term as well as its phenomenological
implications.

\begin{table}[ht]
\caption{\label{tab1}
Contributions of successive orders of in $\alpha_s$   to $\delta \, r^{(q)kl}$
for the value of  $\alpha_s(M_{\tau}) = .334$, and normalized 
to the value of $\delta \, r^{(q)kl}$ in the Born approximation.
First five lines:  fixed order perturbation theory , second five lines: contour improved version,
with RG improvement  for the Adler function  $D^{(q)}$ ; last five lines: contour improved
version with RG summation made for the polarization operator  $\Pi^{(q)}$.
}
\begin{ruledtabular}
\begin{tabular*}{\hsize}{l@{\extracolsep{0ptplus1fil}}c@{\extracolsep{0ptplus1fil}}r}
$(kl) $ & Perturbative series
          \\
\colrule
(0,0)  &  $1\ + \ 0.425\ + \ 0.283\ + \ 0.178\ +  \ 0.0987 = \ 1.98$	 
\\    	  						       	 
(1,0)  &  $1\ + \ 0.532\ + \ 0.458\ + \ 0.423\ +  \ 0.437 = \ 2.85$ 	 
\\    	  						       	 
(2,0)  &  $1\ + \ 0.606\ + \ 0.589\ + \ 0.623\ + \  0.734 = \ 3.55$ 	 
 \\   	  						        	 
(3,0)  &  $1\ + \ 0.663\ + \ 0.695\ + \ 0.793\ + \  1.00  = \ 4.15$ 	 
\\    	  						       	 
(4,0)  &  $1\ + \ 0.708\ + \ 0.784\ + \ 0.943\ + \  \ 1.24 = \ 4.68$    
\\    
\colrule
(0,0) &  $0.753\ + \ 0.214\ + \ 0.065\  - \ 0.0611\ - \ 0.213 = \  0.76$
\\    							      
(1,0) & $0.912\ + \ 0.334\ + \ 0.192\ + \ 0.0675\ - \ 0.0969 = \  1.41$
\\    							      
(2,0) & $1.05\ + \ 0.451\ + \ 0.33\ + \ 0.228\ + \ 0.0802 = \ 2.14$     
 \\   							      
(3,0) & $1.19\ + \ 0.571\ + \ 0.484\ + \ 0.425\ + \ 0.33 = \ 3.$	      
\\    							      
(4,0) & $1.32\ + \ 0.697\ + \ 0.657\ + \ 0.665\ + \ 0.664 = \ 4.01$     
\\
\colrule								 
(0,0) & $0.857\ + \ 0.122\ - \ 0.0090\ - \ 0.158\ - \ 0.355 = \ 0.458$	 
\\     									 
(1,0) & $1.11\ + \ 0.232\ + \ 0.11\  \ -0.0417\ - \ 0.280 = \ 1.13$	 
\\     									 
(2,0) & $1.35\ + \ 0.347\ + \ 0.251\ + \ 0.124\ - \ 0.121 = \ 1.95$	 
 \\    									 
(3,0) & $1.59\ + \ 0.471\ + \ 0.42\ + \ 0.35\ + \ 0.145 = \ 2.97$		 
\\     									 
(4,0) & $1.83\ + \ 0.61\ + \ 0.623\ + \ 0.648\ + \ 0.544 = \ 4.26$         
\\    
\end{tabular*}
\end{ruledtabular}
\end{table}

\section{Calculation and Results}

To compute the real part of $\Pi^{(q)}_2(q^2)$ we proceed as
follows. 
First, using the criterion of irreducibility of Feynman
integrals \cite{Baikov:criterion:00}, the set of irreducible integrals involved in
the problem was constructed.  Second, the
coefficients multiplying  these integrals were calculated as series in
the $1/D\rightarrow0$ expansion. 
Third, the exact answer, i.e.  a rational function of $D$, 
was reconstructed from this expansion.
Our results read:
\bea
\lefteqn{\Pi^{(q)}_{2,us} =
 -4 -\frac{28}{3}\, a_s 
+ a_s^2 \left\{
 -\frac{13981}{108} - \frac{646}{27} \, \zeta_{3}  
+\frac{2080}{27}
 \,\zeta_{5}\right\}
\nonumber
} \ \ \ \ \ \ \  
\\
&{+}& a_s^3 \left\{
 -\frac{2092745}{1296}-\frac{14713}{162} \,\zeta_{3} 
-122 \,\zeta_3^2
+10 \,\zeta_{4} 
\right. 
\nonumber 
\\ &{}& 
\left.
\  \ \ \ \ \ 
+\frac{41065}{27} \,\zeta_{5}  
-\frac{79835}{162}\, \zeta_7
\right\}
\label{Pi2m2us:exact}
\\
&{}& \hspace{-.6cm}= -4  \left(  1 +2.333\, a_s +
19.58\  a_s^2 
+   202.309\  a_s^3
 \right)  {},
\nonumber
\end{eqnarray}

\bea
\Pi^{(q)}_{2,ud} &=&
 a_s^2 \left\{
\frac{128}{9} - \frac{32}{3} \, \zeta_{3}  
\right\}
\label{Pi2m2ud:exact}
\\
&+& a_s^3 \left\{
\frac{6392}{27} - \frac{4496}{27} \,\zeta_{3} 
-
16 \,\zeta_3^2
+\frac{320}{27}\,\zeta_{5} 
\right\}
\nonumber
\\
&=& 
-4  \left(  -0.35\, a_s^2 
-6.437\  a_s^3 
 \right) 
\nonumber
{}.
\end{eqnarray}

\section{Phenomenology}

First of all, it is instructive to compare the exact result for the ${\cal
O}(\als^3)$ contribution to (\ref{Pi2m2us:exact}) with the recently obtained
predictions (\cite{ChBK:tau:as4nf2}) 
based on the  optimization schemes as PMS and FAC:
\beq
k^{(q)3}_{2,us} =  202.309\,(\mbox{exact}),  
\ \ \ 201 \,(\mbox{PMS}), \ \ \ 199 \,(\mbox{FAC})
\label{Kq3_2:predcitions}
{}.
\eeq
This astonishingly good agreement (a similar phenomenon   has been observed for 
the prediction of the $O(\alpha_s^3)$ term for the correlator of the diagonal currents
\cite{ChBK:LL04}  can  be considered as  a strong argument to repeat
the procedure  and predict, starting from the now completely known
$k_2^{(q)3}$,  the corresponding result for one loop
more, that is for $k_2^{(q)4}$.  To be
definite, we use  the PMS predictions (again for $n_f=3$; the FAC
result is  very similar)
\beq
k^{(q)4}_{2,us} =  2276 \pm 200  \  \ \mbox{and}  \ \ 
k^{(q)4}_{2,us} - k^{(q)4}_{2,ud} = 2378 \pm 200 
{}.
\eeq
It is, of course, difficult to estimate 
uncertainty in the above predictions; however, the simple
comparison with  eq.~(\ref{Kq3_2:predcitions}) clearly demonstrates that an error of about
10\% should be considered quite conservative.

Apart of evaluations using fixed order perturbation theory two formally equivalent
versions of the contour improved procedure can be found in the
literature. The first \cite{Chetyrkin:1998ej} 
is based directly on the integration of the
polarization function $\Pi^{(q)}_2$, the second \cite{Pich:1998yn} 
is based on the
integration of the Adler function 
$
D^{(q)}_2 \equiv s\frac{d}{d  s}\Pi^{(q)}_2 
$
and is obtained from the first one by partial
integration.  
Partial integration and renormalization group
improvement do not commute as long as finite orders are
considered. Since the second procedure moves part  of the lower
order input to higher orders (contrary to the spirit of CIPT) and
since, furthermore, the first procedure leads to a somewhat more stable
perturbation series \footnote{At least this is true for small 
values of $\alpha_s(M_{\tau})$; for
realistic higher value $\alpha_s(M_{\tau}) = .334$ both series start to oscillate.}, 
we consider the first of the two choices
as preferable. 
The three options for the perturbative series are displayed in 
Table~\ref{tab1}. As a consequence of the large value of $\alpha_s$ and the rapidly
growing coefficients of the perturbative series it seems at first
glance difficult to consider any of the theoretical predictions as
truly preferable. Nevertheless, the results for moments $(2,0),
(3,0)$ and $(4,0)$ are at least in plausible agreement among the three
methods and exhibit acceptably decreasing subsequent terms. Since
these moments are also relatively most precise, as far as experiment
is concerned, they will be used in the subsequent analysis.
The phenomenological analysis will be based on the most recent
evaluation  \cite{Czarnecki:2004cw} of $|V_{us}|$.
For the phenomenological description of the contribution due to 
$r^{(0)kl}$  we adopt the analysis presented in \cite{Gamiz:2004ar}
for the second version of contour improvement.

The results for $m_s$, derived from different moments and different
ways of implementing the contour improvement procedure are shown in
Table~\ref{tab3}. (Details about the error estimates and the
corresponding analysis will be given elsewhere.) The main difference,
compared to the previous analysis, is a downward shift of $m_s$ by
about 5 MeV from the inclusion of the $\als^4$ terms and an upward
shift by as much as 20 MeV from the new input for
$|V_{us}|$. The ambiguity for the determined value of $m_s$, including the
newly computed $\als^3$ term, and the estimate for the $\als^4$ term
is shown in Table~\ref{tab3}. The renormalzation scale
$\mu = \xi M_{\tau}$ is allowed to vary between $\xi = 1 - 1.5 $.
Values of $\mu$ lower than $M_{\tau}$ lead to a blow up of $\als$ and
destabilize the result.  By ``others'' we mean all uncertainties
(added in quadrature) of the input parameters different from the
${\cal O}(\as^3)$ (or ${\cal O}(\as^4)$) terms in the perturbative
contribution. They include experimental errors in the moments as
reported in
\cite{Abbiendi:2004xa}, in   $|V_{us}| = 0.2259(23)$ from   \cite{Czarnecki:2004cw}             
as well as uncertainties in the                         
parameters related to construction of  the subtracted longitudinal                             
part.  We have computed the   uncertainties using numbers                            
from \cite{Gamiz:2004ar}.

\begin{table}[ht]                                                                       
\renewcommand{\arraystretch}{1.2}                                                      
\begin{center}                                                                         
\begin{tabular}{crrrr}                                                                 
\hline                                                                                 
Parameter   & Value & \quad(2,0)\quad & \quad(3,0)\quad & \quad(4,0)  
\\
\hline  
$m_s({\cal O}(\as^3),\mbox{exact})$ &    &   123.  &   103.  &   88.        
 \\ 
\hline                             
${\cal O}(\as^3)$ & $\asthree$ &         ${}^{-3.5}_{3.9}$  &   ${}^{-5.8}_{7.}$  &   ${}^{-6.8}_{8.9}$      
 \\ 
\hline                             
$\xi$ & ${}^{1.5}_{1}$ &               ${}^{-1.4}_{0}$  &   ${}^{4.5}_{0}$  &   ${}^{7.5}_{0}$     
\\  
\hline                             
$\alpha_s(M_{\tau})$ & $.334 \pm 0.022$ &     ${}^{3.8}_{-1.4}$  &   ${}^{0.44}_{1.2}$  &   ${}^{-1.5}_{2.8}$  
\\                 
\hline                                       
others \cite{Abbiendi:2004xa,Gamiz:2004ar,Czarnecki:2004cw}& &${}^{+22.3}_{-25.7}$ & ${}^{+17.}_{-19.1}$ & ${}^{+14.3}_{-15.8}$
\\                        
\hline                                                                                 
Total    &  & ${}^{+23.}_{-26.}$ & ${}^{+19.}_{-20.}$ & ${}^{+18.6}_{-17.3}$
\\  
\hline
$m_s({\cal O}(\as^4),\mbox{PMS})$ &    & 127.  &   100.  &   82.4      
 \\ 
\hline                             
${\cal O}(\as^4)$ & $\asfour$ &             ${}^{3.9}_{-3.6}$  &   ${}^{-2.3}_{2.4}$  &   ${}^{-4.6}_{5.6}$      
 \\ 
\hline                             
$\xi$ & ${}^{1.5}_{1}$ &        ${}^{-9.1}_{0}$  &   ${}^{-2.}_{0}$  &   ${}^{4.4}_{0}$          
\\  
\hline                             
$\alpha_s(M_{\tau})$ & $.334 \pm 0.022$ &  ${}^{13.}_{-5.6}$  &   ${}^{4.3}_{-0.71}$  &   ${}^{0.24}_{2.1}$   
\\                 
\hline                                       
others \cite{Abbiendi:2004xa,Gamiz:2004ar,Czarnecki:2004cw}& &${}^{+22.9}_{-26.4}$ & ${}^{+16.7}_{-18.8}$ & ${}^{+13.8}_{-15.3}$
\\                        
\hline                                                                                 
Total    &  & ${}^{+26.7}_{-28.6}$ & ${}^{+17.5}_{-19.}$ & ${}^{+15.7}_{-16.}$
\\  
\hline
\end{tabular}                                                                          
\end{center}                                                                           
\caption{
Result for $m_s$, derived from different levels of approximation, based on contour
improvement from  \cite{ChBK:tau:as4nf2} and a list of different contributions to the associated  error.
\label{tab3}}
\end{table}

From Table \ref{tab1} we find that the inclusion of 
the $\als^4$  term leads to a better agreement between predictions based 
on two different methods of implementing of the ``contour improvement''
approach for the third and the fourth moments. As the former moment also shows
smaller theoretical error involved we choose it 
to derive our final result  for $m_s$ which will be  given below.

In total we find 
\beq
m_s(M_{\tau}) = 100 
+ \left({}^{+5.}_{-3.} \right)_{\mbox{theo}} \ \
+  \left( {}^{+17.}_{-19.}  \right)_{\mbox{rest}}
\ \mbox{MeV}
\label{our:fin}
\ 
{}.
\eeq 
If one compares (\ref{our:fin}) to the ${\cal O}(\als^3)$  result of \cite{Gamiz:2004ar} 
(corrected for a different $|V_{us}|$)  
$m_s(M_{\tau}) = 106\ \mbox{MeV} $ (with larger theoretical and
identical remaining errors) one sees 
an  essential but still not too large sensitivity to the
$\alpha_s^4$ contribution.  In fact, for the third moment the shift in
$m_s$ due inclusion of the $\alpha_s^4$ term (-2.5 MeV) is about a third of 
of the corresponding change (-7 MeV) due to the  $\alpha_s^3$
contribution.  Thus,  the purely theoretical uncertainty from not
yet computed higher orders could be estimated as about 3 MeV.
Unfortunately, one can hardly hope that the error from not yet
computed higher orders in $\alpha_s$ could be reduced further by means
of a direct calculation in any foreseeable future.

The authors are grateful to A.~Pivovarov and K.~Maltman
for   very useful discussions.  This work was supported by the
the Deutsche Forschungsgemeinschaft in the Sonderforschungsbereich/Transregio 
SFB/TR-9 ``Computational Particle Physics'',  by INTAS (grant
03-51-4007), by RFBR (grant 03-02-17177) and by Volkswagen Foundation
(grant I/77788).


\begin{thebibliography}{36}
\expandafter\ifx\csname natexlab\endcsname\relax\def\natexlab#1{#1}\fi
\expandafter\ifx\csname bibnamefont\endcsname\relax
  \def\bibnamefont#1{#1}\fi
\expandafter\ifx\csname bibfnamefont\endcsname\relax
  \def\bibfnamefont#1{#1}\fi
\expandafter\ifx\csname citenamefont\endcsname\relax
  \def\citenamefont#1{#1}\fi
\expandafter\ifx\csname url\endcsname\relax
  \def\url#1{\texttt{#1}}\fi
\expandafter\ifx\csname urlprefix\endcsname\relax\def\urlprefix{URL }\fi
\providecommand{\bibinfo}[2]{#2}
\providecommand{\eprint}[2][]{\url{#2}}

\bibitem[{\citenamefont{Barate et~al.}(1999)}]{Barate:1999hj}
\bibinfo{author}{\bibfnamefont{R.}~\bibnamefont{Barate}} \bibnamefont{et~al.}
  (\bibinfo{collaboration}{ALEPH}), \bibinfo{journal}{Eur. Phys. J.}
  \textbf{\bibinfo{volume}{C11}}, \bibinfo{pages}{599} (\bibinfo{year}{1999}),
  \eprint{hep-ex/9903015}.

\bibitem[{\citenamefont{Abbiendi et~al.}(2004)}]{Abbiendi:2004xa}
\bibinfo{author}{\bibfnamefont{G.}~\bibnamefont{Abbiendi}} \bibnamefont{et~al.}
  (\bibinfo{collaboration}{OPAL}), \bibinfo{journal}{Eur. Phys. J.}
  \textbf{\bibinfo{volume}{C35}}, \bibinfo{pages}{437} (\bibinfo{year}{2004}),
  \eprint{hep-ex/0406007}.

\bibitem[{\citenamefont{Chetyrkin and Kwiatkowski}(1993)}]{Chetyrkin:1993hi}
\bibinfo{author}{\bibfnamefont{K.~G.} \bibnamefont{Chetyrkin}}
  \bibnamefont{and}
  \bibinfo{author}{\bibfnamefont{A.}~\bibnamefont{Kwiatkowski}},
  \bibinfo{journal}{Z. Phys.} \textbf{\bibinfo{volume}{C59}},
  \bibinfo{pages}{525} (\bibinfo{year}{1993}), \eprint{hep-ph/9805232}.


\bibitem[{\citenamefont{Maltman}(1998)}]{Maltman:1998qz}
\bibinfo{author}{\bibfnamefont{K.}~\bibnamefont{Maltman}},
  \bibinfo{journal}{Phys. Rev.} \textbf{\bibinfo{volume}{D58}},
  \bibinfo{pages}{093015} (\bibinfo{year}{1998}), \eprint{hep-ph/9804298}.



\bibitem[{\citenamefont{Chetyrkin et~al.}(1998)\citenamefont{Chetyrkin,
  K{\"u}hn, and Pivovarov}}]{Chetyrkin:1998ej}
\bibinfo{author}{\bibfnamefont{K.~G.} \bibnamefont{Chetyrkin}},
  \bibinfo{author}{\bibfnamefont{J.~H.} \bibnamefont{K{\"u}hn}},
  \bibnamefont{and} \bibinfo{author}{\bibfnamefont{A.~A.}
  \bibnamefont{Pivovarov}}, \bibinfo{journal}{Nucl. Phys.}
  \textbf{\bibinfo{volume}{B533}}, \bibinfo{pages}{473} (\bibinfo{year}{1998}),
  \eprint{hep-ph/9805335}.

\bibitem[{\citenamefont{Pich and Prades}(1998)}]{Pich:1998yn}
\bibinfo{author}{\bibfnamefont{A.}~\bibnamefont{Pich}} \bibnamefont{and}
  \bibinfo{author}{\bibfnamefont{J.}~\bibnamefont{Prades}},
  \bibinfo{journal}{JHEP} \textbf{\bibinfo{volume}{06}}, \bibinfo{pages}{013}
  (\bibinfo{year}{1998}), \eprint{hep-ph/9804462};
  \bibinfo{journal}{JHEP} \textbf{\bibinfo{volume}{10}}, \bibinfo{pages}{004}
  (\bibinfo{year}{1999}), \eprint{hep-ph/9909244}.






\bibitem[{\citenamefont{Korner et~al.}(2001)\citenamefont{Korner, Krajewski,
  and Pivovarov}}]{Korner:2000wd}
\bibinfo{author}{\bibfnamefont{J.~G.} \bibnamefont{Korner}},
  \bibinfo{author}{\bibfnamefont{F.}~\bibnamefont{Krajewski}},
  \bibnamefont{and} \bibinfo{author}{\bibfnamefont{A.~A.}
  \bibnamefont{Pivovarov}}, \bibinfo{journal}{Eur. Phys. J.}
  \textbf{\bibinfo{volume}{C20}}, \bibinfo{pages}{259} (\bibinfo{year}{2001}),
  \eprint{hep-ph/0003165}.

\bibitem[{\citenamefont{Chen et~al.}(2001)}]{Chen:2001qf}
\bibinfo{author}{\bibfnamefont{S.}~\bibnamefont{Chen}} \bibnamefont{et~al.},
  \bibinfo{journal}{Eur. Phys. J.} \textbf{\bibinfo{volume}{C22}},
  \bibinfo{pages}{31} (\bibinfo{year}{2001}), \eprint{hep-ph/0105253}.

\bibitem[{\citenamefont{Maltman}(2002)}]{Maltman:2002wb}
\bibinfo{author}{\bibfnamefont{K.}~\bibnamefont{Maltman}},
  \bibinfo{journal}{eConf} \textbf{\bibinfo{volume}{C0209101}},
  \bibinfo{pages}{WE05} (\bibinfo{year}{2002}), \eprint{hep-ph/0209091}.

\bibitem[{\citenamefont{Kuhn and Mirkes}(1992)}]{Kuhn:1992nz}
\bibinfo{author}{\bibfnamefont{J.~H.} \bibnamefont{Kuhn}} \bibnamefont{and}
  \bibinfo{author}{\bibfnamefont{E.}~\bibnamefont{Mirkes}},
  \bibinfo{journal}{Z. Phys.} \textbf{\bibinfo{volume}{C56}},
  \bibinfo{pages}{661} (\bibinfo{year}{1992}).

\bibitem[{\citenamefont{Braaten et~al.}(1992)\citenamefont{Braaten, Narison,
  and Pich}}]{Braaten:1991qm}
\bibinfo{author}{\bibfnamefont{E.}~\bibnamefont{Braaten}},
  \bibinfo{author}{\bibfnamefont{S.}~\bibnamefont{Narison}}, \bibnamefont{and}
  \bibinfo{author}{\bibfnamefont{A.}~\bibnamefont{Pich}},
  \bibinfo{journal}{Nucl. Phys.} \textbf{\bibinfo{volume}{B373}},
  \bibinfo{pages}{581} (\bibinfo{year}{1992}).

\bibitem[{\citenamefont{Stevenson}(1981)}]{Stevenson:1981vj}
\bibinfo{author}{\bibfnamefont{P.~M.} \bibnamefont{Stevenson}},
  \bibinfo{journal}{Phys. Rev.} \textbf{\bibinfo{volume}{D23}},
  \bibinfo{pages}{2916} (\bibinfo{year}{1981}).

\bibitem[{\citenamefont{Grunberg}(1984)}]{Grunberg:1984fw}
\bibinfo{author}{\bibfnamefont{G.}~\bibnamefont{Grunberg}},
  \bibinfo{journal}{Phys. Rev.} \textbf{\bibinfo{volume}{D29}},
  \bibinfo{pages}{2315} (\bibinfo{year}{1984}).

\bibitem[{\citenamefont{Baikov}(1996)}]{Baikov:tadpoles:96}
\bibinfo{author}{\bibfnamefont{P.~A.} \bibnamefont{Baikov}},
  \bibinfo{journal}{Phys. Lett.} \textbf{\bibinfo{volume}{B385}},
  \bibinfo{pages}{404} (\bibinfo{year}{1996}), \eprint{hep-ph/9603267};
 \bibinfo{journal}{Nucl. Instrum. Meth.} \textbf{\bibinfo{volume}{A389}},
  \bibinfo{pages}{347} (\bibinfo{year}{1997}), \eprint{hep-ph/9611449}.



\bibitem[{\citenamefont{Baikov and Steinhauser}(1998)}]{BkvSthr:tadples:98}
\bibinfo{author}{\bibfnamefont{P.~A.} \bibnamefont{Baikov}} \bibnamefont{and}
  \bibinfo{author}{\bibfnamefont{M.}~\bibnamefont{Steinhauser}},
  \bibinfo{journal}{Comput. Phys. Commun.} \textbf{\bibinfo{volume}{115}},
  \bibinfo{pages}{161} (\bibinfo{year}{1998}), \eprint{hep-ph/9802429}.

\bibitem[{\citenamefont{Baikov}(2000)}]{Baikov:criterion:00}
\bibinfo{author}{\bibfnamefont{P.~A.} \bibnamefont{Baikov}},
  \bibinfo{journal}{Phys. Lett.} \textbf{\bibinfo{volume}{B474}},
  \bibinfo{pages}{385} (\bibinfo{year}{2000}), \eprint{hep-ph/9912421};
\bibinfo{journal}{Nucl. Phys. Proc. Suppl.} \textbf{\bibinfo{volume}{116}},
  \bibinfo{pages}{378} (\bibinfo{year}{2003}).


\bibitem[{\citenamefont{Vermaseren}(2000)}]{Vermaseren:2000nd}
\bibinfo{author}{\bibfnamefont{J.~A.~M.} \bibnamefont{Vermaseren}}
  (\bibinfo{year}{2000}), \eprint{math-ph/0010025}.


%

\bibitem[{\citenamefont{Pivovarov}(1992)}]{Pivovarov:1991bi}
\bibinfo{author}{\bibfnamefont{A.~A.} \bibnamefont{Pivovarov}},
  \bibinfo{journal}{Nuovo Cim.} \textbf{\bibinfo{volume}{A105}},
  \bibinfo{pages}{813} (\bibinfo{year}{1992}{\natexlab{a}});
\bibinfo{journal}{Z. Phys.} \textbf{\bibinfo{volume}{C53}},
 \bibinfo{pages}{461} (\bibinfo{year}{1992}{\natexlab{b}}).


%
%


\bibitem[{\citenamefont{Le~Diberder and
  Pich}(1992{\natexlab{a}})}]{LeDiberder:1992te}
\bibinfo{author}{\bibfnamefont{F.}~\bibnamefont{Le~Diberder}} \bibnamefont{and}
  \bibinfo{author}{\bibfnamefont{A.}~\bibnamefont{Pich}},
  \bibinfo{journal}{Phys. Lett.} \textbf{\bibinfo{volume}{B286}},
 \bibinfo{pages}{147} (\bibinfo{year}{1992}{\natexlab{a}});
\bibinfo{journal}{Phys. Lett.} \textbf{\bibinfo{volume}{B289}},
 \bibinfo{pages}{165} (\bibinfo{year}{1992}{\natexlab{b}}).





\bibitem[{\citenamefont{Barate et~al.}(1998)}]{Barate:1998uf}
\bibinfo{author}{\bibfnamefont{R.}~\bibnamefont{Barate}} \bibnamefont{et~al.}
  (\bibinfo{collaboration}{ALEPH}), \bibinfo{journal}{Eur. Phys. J.}
  \textbf{\bibinfo{volume}{C4}}, \bibinfo{pages}{409} (\bibinfo{year}{1998}).

\bibitem[{\citenamefont{Ackerstaff et~al.}(1999)}]{Ackerstaff:1998yj}
\bibinfo{author}{\bibfnamefont{K.}~\bibnamefont{Ackerstaff}}
  \bibnamefont{et~al.} (\bibinfo{collaboration}{OPAL}), \bibinfo{journal}{Eur.
  Phys. J.} \textbf{\bibinfo{volume}{C7}}, \bibinfo{pages}{571}
  (\bibinfo{year}{1999}), \eprint{hep-ex/9808019}.

\bibitem[{\citenamefont{Chetyrkin et~al.}(1996)\citenamefont{Chetyrkin, Kuhn,
  and Kwiatkowski}}]{Chetyrkin:1996ia}
\bibinfo{author}{\bibfnamefont{K.~G.} \bibnamefont{Chetyrkin}},
  \bibinfo{author}{\bibfnamefont{J.~H.} \bibnamefont{Kuhn}}, \bibnamefont{and}
  \bibinfo{author}{\bibfnamefont{A.}~\bibnamefont{Kwiatkowski}},
  \bibinfo{journal}{Phys. Rept.} \textbf{\bibinfo{volume}{277}},
  \bibinfo{pages}{189} (\bibinfo{year}{1996}).

\bibitem[{\citenamefont{{The LEP WW Working Group,
  LEPEWWG/2003-02}}(2003)}]{EWWG:2003_v3}
\bibinfo{author}{\bibnamefont{{The LEP WW Working Group, LEPEWWG/2003-02}}}
  (\bibinfo{year}{2003}), \eprint{hep-ex/0312023}.

\bibitem[{\citenamefont{Marciano and Sirlin}(1988)}]{Marciano:1988vm}
\bibinfo{author}{\bibfnamefont{W.~J.} \bibnamefont{Marciano}} \bibnamefont{and}
  \bibinfo{author}{\bibfnamefont{A.}~\bibnamefont{Sirlin}},
  \bibinfo{journal}{Phys. Rev. Lett.} \textbf{\bibinfo{volume}{61}},
  \bibinfo{pages}{1815} (\bibinfo{year}{1988}).

\bibitem[{\citenamefont{Braaten and Li}(1990)}]{Braaten:1990ef}
\bibinfo{author}{\bibfnamefont{E.}~\bibnamefont{Braaten}} \bibnamefont{and}
  \bibinfo{author}{\bibfnamefont{C.-S.} \bibnamefont{Li}},
  \bibinfo{journal}{Phys. Rev.} \textbf{\bibinfo{volume}{D42}},
  \bibinfo{pages}{3888} (\bibinfo{year}{1990}).


\bibitem[{\citenamefont{Kambor and Maltman}(2000)}]{Kambor:2000dj}
\bibinfo{author}{\bibfnamefont{J.}~\bibnamefont{Kambor}} \bibnamefont{and}
  \bibinfo{author}{\bibfnamefont{K.}~\bibnamefont{Maltman}},
  \bibinfo{journal}{Phys. Rev.} \textbf{\bibinfo{volume}{D62}},
  \bibinfo{pages}{093023} (\bibinfo{year}{2000}), \eprint{hep-ph/0005156};
 \bibinfo{journal}{Phys. Rev.} \textbf{\bibinfo{volume}{D64}},
  \bibinfo{pages}{093014} (\bibinfo{year}{2001}), \eprint{hep-ph/0107187}.


\bibitem[{\citenamefont{Gamiz et~al.}(2003)\citenamefont{Gamiz, Jamin, Pich,
  Prades, and Schwab}}]{Gamiz:2002nu}
\bibinfo{author}{\bibfnamefont{E.}~\bibnamefont{Gamiz}},
  \bibinfo{author}{\bibfnamefont{M.}~\bibnamefont{Jamin}},
  \bibinfo{author}{\bibfnamefont{A.}~\bibnamefont{Pich}},
  \bibinfo{author}{\bibfnamefont{J.}~\bibnamefont{Prades}}, \bibnamefont{and}
  \bibinfo{author}{\bibfnamefont{F.}~\bibnamefont{Schwab}},
  \bibinfo{journal}{JHEP} \textbf{\bibinfo{volume}{01}}, \bibinfo{pages}{060}
  (\bibinfo{year}{2003}), \eprint{hep-ph/0212230}.




\bibitem[{\citenamefont{Gamiz et~al.}(2005)\citenamefont{Gamiz, Jamin, Pich,
  Prades, and Schwab}}]{Gamiz:2004ar}
\bibinfo{author}{\bibfnamefont{E.}~\bibnamefont{Gamiz}},
  \bibinfo{author}{\bibfnamefont{M.}~\bibnamefont{Jamin}},
  \bibinfo{author}{\bibfnamefont{A.}~\bibnamefont{Pich}},
  \bibinfo{author}{\bibfnamefont{J.}~\bibnamefont{Prades}}, \bibnamefont{and}
  \bibinfo{author}{\bibfnamefont{F.}~\bibnamefont{Schwab}},
  \bibinfo{journal}{Phys. Rev. Lett.} \textbf{\bibinfo{volume}{94}},
  \bibinfo{pages}{011803} (\bibinfo{year}{2005}), \eprint{hep-ph/0408044}.



\bibitem[{\citenamefont{Gorbunov and Pivovarov}(2005)}]{Gorbunov:2004wy}
\bibinfo{author}{\bibfnamefont{D.~S.} \bibnamefont{Gorbunov}} \bibnamefont{and}
  \bibinfo{author}{\bibfnamefont{A.~A.} \bibnamefont{Pivovarov}},
  \bibinfo{journal}{Phys. Rev.} \textbf{\bibinfo{volume}{D71}},
  \bibinfo{pages}{013002} (\bibinfo{year}{2005}), \eprint{hep-ph/0410196}.



\bibitem[{\citenamefont{Gorishnii et~al.}(1991)\citenamefont{Gorishnii, Kataev,
  and Larin}}]{Gorishnii:1991vf}
\bibinfo{author}{\bibfnamefont{S.~G.} \bibnamefont{Gorishnii}},
  \bibinfo{author}{\bibfnamefont{A.~L.} \bibnamefont{Kataev}},
  \bibnamefont{and} \bibinfo{author}{\bibfnamefont{S.~A.} \bibnamefont{Larin}},
  \bibinfo{journal}{Phys. Lett.} \textbf{\bibinfo{volume}{B259}},
  \bibinfo{pages}{144} (\bibinfo{year}{1991}).



\bibitem[{\citenamefont{Baikov et~al.}(2003)\citenamefont{Baikov, Chetyrkin,
  and K{\"u}hn}}]{ChBK:tau:as4nf2}
\bibinfo{author}{\bibfnamefont{P.~A.} \bibnamefont{Baikov}},
  \bibinfo{author}{\bibfnamefont{K.~G.} \bibnamefont{Chetyrkin}},
  \bibnamefont{and} \bibinfo{author}{\bibfnamefont{J.~H.}
  \bibnamefont{K{\"u}hn}}, \bibinfo{journal}{Phys. Rev.}
  \textbf{\bibinfo{volume}{D67}}, \bibinfo{pages}{074026}
  (\bibinfo{year}{2003}), \eprint{hep-ph/0212299}.

\bibitem[{\citenamefont{Baikov et~al.}(2004)\citenamefont{Baikov, Chetyrkin,
  and K{\"u}hn}}]{ChBK:LL04}
\bibinfo{author}{\bibfnamefont{P.~A.} \bibnamefont{Baikov}},
  \bibinfo{author}{\bibfnamefont{K.~G.} \bibnamefont{Chetyrkin}},
  \bibnamefont{and} \bibinfo{author}{\bibfnamefont{J.~H.}
  \bibnamefont{K{\"u}hn}}, \bibinfo{journal}{Nucl. Phys. Proc. Suppl.}
  \textbf{\bibinfo{volume}{135}}, \bibinfo{pages}{243} (\bibinfo{year}{2004}).


\bibitem[{\citenamefont{Czarnecki et~al.}(2004)\citenamefont{Czarnecki,
  Marciano, and Sirlin}}]{Czarnecki:2004cw}
\bibinfo{author}{\bibfnamefont{A.}~\bibnamefont{Czarnecki}},
  \bibinfo{author}{\bibfnamefont{W.~J.} \bibnamefont{Marciano}},
  \bibnamefont{and} \bibinfo{author}{\bibfnamefont{A.}~\bibnamefont{Sirlin}},
  \bibinfo{journal}{Phys. Rev.} \textbf{\bibinfo{volume}{D70}},
  \bibinfo{pages}{093006} (\bibinfo{year}{2004}), \eprint{hep-ph/0406324}.






\end{thebibliography}
\end{document}